\documentclass[11pt]{article}
\pdfoutput=1
\usepackage{geometry}                
\usepackage{graphicx}
\usepackage{amssymb}
\usepackage{amsmath}
 \let\t=\tau

\let\w=\omega

\def\ba{\begin{array}}
\def\ea{\end{array}}

\def\dalemb#1#2{{\vbox{\hrule height .#2pt
        \hbox{\vrule width.#2pt height#1pt \kern#1pt
                \vrule width.#2pt}
        \hrule height.#2pt}}}

 \newcommand{\be}{\begin{equation}}
\newcommand{\ee}{\end{equation}}
 \newcommand{\ben}{\begin{equation*}}
\newcommand{\een}{\end{equation*}}
\newcommand{\bea}{\begin{eqnarray}}
\newcommand{\eea}{\end{eqnarray}}
\newcommand{\bean}{\begin{eqnarray*}}
\newcommand{\eean}{\end{eqnarray*}}

\newcommand{\zb}{\bar{z}}
\newcommand{\phib}{\bar{\phi}}
\newcommand{\wb}{\bar{w}}

\begin{document}

\begin{flushright}
NSF-KITP-08-75 \\
arXiv:0805.3898 [hep-th]
\end{flushright}

\begin{center}
\vspace{1cm} { \LARGE {\bf Pseudogap and time reversal breaking
in a holographic superconductor}}

\vspace{1.1cm}

Matthew M.
Roberts$^\sharp$ and Sean A. Hartnoll$^\flat$

\vspace{0.8cm}

{\it $^\sharp$ Department of Physics, University of California\\
     Santa Barbara, CA 93106-9530, USA }

\vspace{0.5cm}

{\it $^\flat$ KITP, University of California\\
     Santa Barbara, CA 93106-4030, USA }

\vspace{0.8cm}

{\tt matt@physics.ucsb.edu, hartnoll@kitp.ucsb.edu} \\

\vspace{1.2cm}

\end{center}

\begin{abstract}

\noindent Classical $SU(2)$ Yang-Mills theory in 3+1 dimensional
anti-de Sitter space is known to provide a holographic dual to a
2+1 system that undergoes a superconducting phase transition. We
study the electrical conductivity and spectral density of an
isotropic superconducting phase. We show that the theory exhibits
a pseudogap at low temperatures and a nonzero Hall conductivity.
The Hall conductivity is possible because of spontaneous breaking
of time reversal symmetry.

\end{abstract}

\pagebreak
\setcounter{page}{1}

\section{Introduction}

The first successful microscopic theory of superconductivity, BCS
theory \cite{Bardeen:1957mv}, was developed over fifty years ago
and correctly describes the superconducting phenomenology of a
large number of metals and alloys \cite{parks}. The essence of
superconductivity is the spontaneous breaking at low temperatures
of a $U(1)$ symmetry due to a charged condensate. In BCS theory,
the condensate is a Cooper pair of electrons, bound together by
lattice vibrations, or phonons.

It has been appreciated for some time that materials of
significant theoretical and practical interest, such as the heavy
fermion compounds \cite{heavy, heavy2} or the high $T_c$ cuprates
\cite{hightc}, require new theoretical input. For these materials,
neither the pairing mechanism, leading to the charged condensate,
nor the properties of the superconducting state itself are those
of BCS theory. Furthermore, there are indications that the
relevant new physics is strongly coupled, requiring a departure
from the quasiparticle paradigm of Fermi liquid theory
\cite{heavy, hightc}.

Our hope is that a solvable model of a strongly coupled system
undergoing a superconducting phase transition might help the
development of new theories of superconductivity.
It has recently been shown that the
AdS/CFT correspondence \cite{Maldacena:1997re} can indeed provide
models of strongly interacting superconductors in which
calculations can be performed from first principles
\cite{Gubser:2005ih, Gubser:2008px, Hartnoll:2008vx, Gubser:2008zu}.
These recent works are part of a wider program of applying the
AdS/CFT correspondence to condensed matter systems
\cite{Herzog:2007ij, Hartnoll:2007ih, Hartnoll:2007ip,
Hartnoll:2008hs, Son:2008ye, Balasubramanian:2008dm}. The
philosophy is that even if the underlying microscopic descriptions
of theories with AdS duals are likely quite different to those
arising in materials of experimental interest, aspects of the
strongly coupled dynamics and kinematics may be universal.
Kinematically speaking, theories with AdS duals are quantum
critical \cite{sachdev}. The superconductors described to date
within the AdS/CFT framework are quantum critical systems that
undergo a superconducting phase transition as a function of
temperature over chemical potential.

In the recent work \cite{Hartnoll:2008vx}, the electrical
conductivity of a holographic (i.e. AdS/CFT) superconductor was
computed as a function of frequency. A delta function at the
origin, $\w = 0$, due to the Goldstone boson of the broken $U(1)$
symmetry, was followed by a gap $2 \Delta$ in which the
dissipative conductivity vanished at zero temperature. At small
but finite temperature the conductivity in the gap was suppressed
by $e^{- \Delta/T}$. Beyond the gap, a finite spectral density
was observed, exhibiting or not a coherence peak depending on the
details of the system. What was remarkable about these results was
that despite coming from a strongly coupled quantum critical
theory, they are exactly the qualitative features of
conductivity in the superconducting phase that one obtains from
weakly coupled BCS theory \cite{parks}. The main difference, and
apparently only indication of an underlying strong coupling
dynamics, was that \cite{Hartnoll:2008vx} found the zero temperature
gap $2 \Delta/T_c \approx 8.4$ rather than $2 \Delta/T_c \approx 3.5$
for BCS theory.

It is of interest to find and study gravitational duals to
superconductivity with qualitative features that are not those of
BCS theory, but rather of nonconventional superconductors such as
the heavy fermions or the cuprates. In this paper we will study
the electrical conductivity of a holographic superconductor
recently proposed by Gubser \cite{Gubser:2008px}. We show that
this model exhibits two such nonconventional features: a
`pseudogap' rather than a gap at zero temperature and spontaneous
breaking of time reversal invariance.

By `pseudogap' we do not mean the exotic and controversial region
in the normal phase of cuprate superconductors \cite{hightc}.
Rather, we will use the term to denote a well-defined gap in the
dissipative conductivity at low frequencies in which the conductivity
is not identically zero. In nonconventional superconductors, this
pseudogap is due to the fact that the Cooper pairs are not bound
states with zero angular momentum ($l=0$) but rather so-called
$p$-wave ($l=1$) or $d$-wave ($l=2$ spin singlet) states. The gap
above the Fermi surface in these superconductors vanishes at
certain specific directions in momentum space and therefore one
finds (a reduced number of) excitations with arbitrarily low
energy \cite{hightc, heavy2}.

Spontaneous breaking of time reversal invariance has recently been
observed, for instance, in the YBCO high $T_c$ superconductor
\cite{one,two,three,four,five}. The breaking is thought to occur
because the condensing Cooper pairs are not only not $s$-waves but
in fact a complex combination of $d$-waves: $d_{x^2-y^2} + i
d_{xy}$. The simple fact that this is a supposition of T invariant
states with differing phases is sufficient to break T invariance.
Recall that T is an anti-linear operator. One immediate
consequence of breaking time reversal is that it is possible for
these systems to have a Hall conductivity even in the absence of
an external magnetic field \cite{tone, ttwo}. We will review this
fact below.

The layout of this paper is as follows. We will first review the
Einstein-Yang-Mills system that was shown in \cite{Gubser:2008zu}
to be a holographic dual to a theory with a superconducting phase
transition that spontaneously breaks T invariance. We will compute
the electric conductivity and the spectral density in the
superconducting phase and exhibit a pseudogap and a Hall
conductivity.
\vspace{0.2cm}

\noindent {\bf Note:} As this paper was nearing completion, the preprint
\cite{Gubser:2008wv} was posted to the arxiv, with some overlap with this
work. Furthermore, it was pointed out in \cite{Gubser:2008wv} that
the superconducting phase we will be studying is in fact unstable
near $T_c$ to the breaking of rotational invariance. We have
extended and confirmed this stability analysis in section
\ref{sec:stability} below. The instability is in a mode
orthogonal to the modes we study. Therefore our results for the
conductivity should be interesting as a strong coupling
computation of an isotropic pseudogapped superconducting phase
with both Hall and direct conductivities (with no external
magnetic field).

\section{The Einstein-Yang-Mills background}

The gravitational geometry dual to our finite temperature
superconductor will be the planar Schwarzschild-AdS black hole in
3+1 dimensions

\be\label{eq:blackhole}
ds^2=\frac{r^2}{L^2}\Big(-h(r) dt^2+ 2 dz
d\bar{z}\Big)+\frac{L^2}{r^2 h(r)}dr^2
\,,
\ee
where $h(r) = 1-\frac{r_0^3}{r^3}$ and we have introduced the
complex coordinates
\be\label{eq:z}
z = \frac{x + i y}{\sqrt{2}} \,.
\ee
The scale $L$ is the AdS radius and $r_0$ gives the location of
the black hole horizon. The geometry (\ref{eq:blackhole}) is a
solution of the vacuum Einstein equations with a negative
cosmological constant.

On this background we will study an $SU(2)$ gauge theory. This was
suggested to be an interesting dual for a superconductor in
\cite{Gubser:2008zu}, and we shall review the dynamics shortly. The full
Einstein-Yang-Mills theory has the action
\be\label{eq:action}
S = \int d^4x \sqrt{-g} \left[ \frac{1}{2 \kappa^2_4} \left( R + \frac{6}{L^2} \right)
-\frac{1}{4 g^2}F^a_{\mu\nu}F^{a\mu\nu} \right] \,.
\ee
Following \cite{Hartnoll:2008vx, Gubser:2008zu}, we will be
working in the probe limit, in which the $SU(2)$ fields are small
and do not backreact on the metric\footnote{The probe limit is in
fact necessary to obtain a finite DC ($\w=0$) conductivity even in
the normal phase. A translationally invariant system with a finite
charge density will have an infinite DC conductivity unless the
momentum can leak somewhere. In the probe limit, momentum is
leaked to the metric or `glue' sector. See for instance
\cite{Karch:2007pd}.}. Conformal invariance of the Yang-Mills
action means that by rescaling the metric, keeping the Yang-Mills
coupling $g$ fixed, we can always make the probe approximation
consistently. Specifically, we take $L g \gg \kappa_4$. In fact,
by rescaling all the fields we can set $L=1$, which we will
proceed to do. Let us now set up some notation.

The three generators $\tau^a$ of the $SU(2)$ algebra satisfy
$[\tau^b,\tau^c]=\tau^a f^{abc}$. In the standard basis we have
$f^{abc} = \epsilon^{abc}$, with $\epsilon^{123}=1$.
It will be useful for us to work with the combinations
\be
\tau^{\pm} = \frac{\tau^1\pm i\tau^2}{\sqrt{2}} \,,
\ee
as well as $\tau^3$. The field strength is
\be
F^a_{\mu \nu} = \partial_\mu A^a_\nu-\partial_\nu A^a_\mu+
f^{abc}A^b_\mu A^c_\nu \,.
\ee
The equations of motion are
\be
\frac{1}{\sqrt{-g}}\partial_\mu(\sqrt{-g}F^{a\mu\nu})+f^{abc}A^b_\mu
F^{c\mu\nu} = 0 \,.
\ee

Following \cite{Gubser:2008zu} we will identify the $U(1)$
subgroup of $SU(2)$ generated by $\t^3$ to be the electromagnetic
$U(1)$. The W-bosons $A^{\pm}$ are therefore charged fields. The
insight of \cite{Gubser:2008zu} was to note that a sufficiently
large background electric field for the $U(1)$, at fixed
temperature, would cause the W-bosons to condense and hence
trigger superconductivity. Specifically, consider the background
ansatz
\be\label{eq:ansatz}
A= \phi(r)~dt\tau^3+ w(r)~(dz \tau^-+d\zb \tau^+) \,.
\ee
An important feature of this ansatz is that whereas the original
rotational invariance is broken, the system is invariant under a
combined gauge and spatial rotation
\be\label{eq:combined}
z \rightarrow e^{i\theta} z\,, \qquad \tau^\pm \rightarrow e^{\pm
i\theta}\tau^\pm \,.
\ee
Therefore the superconducting phase we are considering is
effectively spatially isotropic. A symmetry that is broken by the
$A^{\pm}_{z,\bar z}(r)$ fields is
time reversal invariance. These gauge potentials lead to a
magnetic field in the bulk. A magnetic fields breaks time
reversal, as can be immediately seen from (for instance) the Lorentz force
law.

Evaluated on this ansatz, the equations of motion become
\bea\label{eq:eom}
\phi''+\frac{2}{r}\phi'-\frac{2 w^2}{r^4 h}\phi & = & 0 \,,
\nonumber \\
w''+(\frac{2}{r}+\frac{h'}{h}) w'+\frac{\phi^2w- h w^3}{r^4 h^2} &
= & 0 \,.
\eea
Under a rescaling $r=r_0
\bar r\,,\phi(r)=r_0\,, \phib(x)\,, w(r)=r_0\wb(x)$, the black hole radius
$r_0$ cancels out of the equations entirely, rendering all
quantities dimensionless. This is a consequence of the dual
theory being a conformal field theory in flat space.
Therefore we will set $r_0 = 1$ in what
follows. The Hawking temperature of the black hole
(\ref{eq:blackhole}) is
\be
T = \frac{3 r_0}{4 \pi} \,.
\ee
We will restore the $r_0$ dependence when we wish to re-express
quantities in terms of the temperature.

There are two solutions to the equations (\ref{eq:eom}). The first
is the AdS-Reissner-Nordstrom black hole which has\footnote{In the following definitions
we have rescaled the boundary charge density and condensate by $\rho,J \to \rho/g^2, J/g^2$,
to eliminate messy factors of $g$. This rescaling does not affect the physical ratio $2\Delta/T_c$.}
\be\label{eq:adsrn}
\phi = \rho (1 - 1/r) \,, \qquad w = 0 \,.
\ee
The scalar potential is required to vanish on the horizon (recall
that we have set $r_0=1$). This ties the chemical potential $\mu$
to equal the charge density $\rho$. The second solution is a hairy
black hole. This solution needs to be found numerically. The
main feature is that $w \neq0 $, but still normalizable at
infinity. Thus at large radius $r$ we require for the scalar
potential
\be
\phi = \mu- \frac{\rho}{r} + \cdots \,,
\ee
and for the charged condensate
\be\label{eq:wlarger}
w = \frac{\langle J \rangle}{\sqrt{2} r} + \cdots \,.
\ee
Here $\langle J \rangle$ is a condensate of the charged operator
dual to $A_z^- \t^- + A_z^+ \t^+$ (up to a sign). It is a component of the global
$SU(2)$ current in the field theory. We have required that there is no source term
in field theory action for the operator $J$, by demanding that the
constant term in the large $r$ expansion of $w$ vanish. From the
field theory point of view, the presence of the $\langle J
\rangle$ condensate in the absence of a source means that time reversal
symmetry is spontaneously broken.

In order to find the hairy black hole solutions, one should
numerically integrate out from the horizon. Near the horizon one
writes $\phi = \phi_1 (r-1) + \cdots$ and $w= w_0 + \cdots$. There
are two constants of integration, $w_0$ and $\phi_1$, and
therefore, upon imposing the falloff (\ref{eq:wlarger}) we are
left with a one parameter family of solutions.

\begin{figure}[h]
\begin{center}
\includegraphics[width=3in]{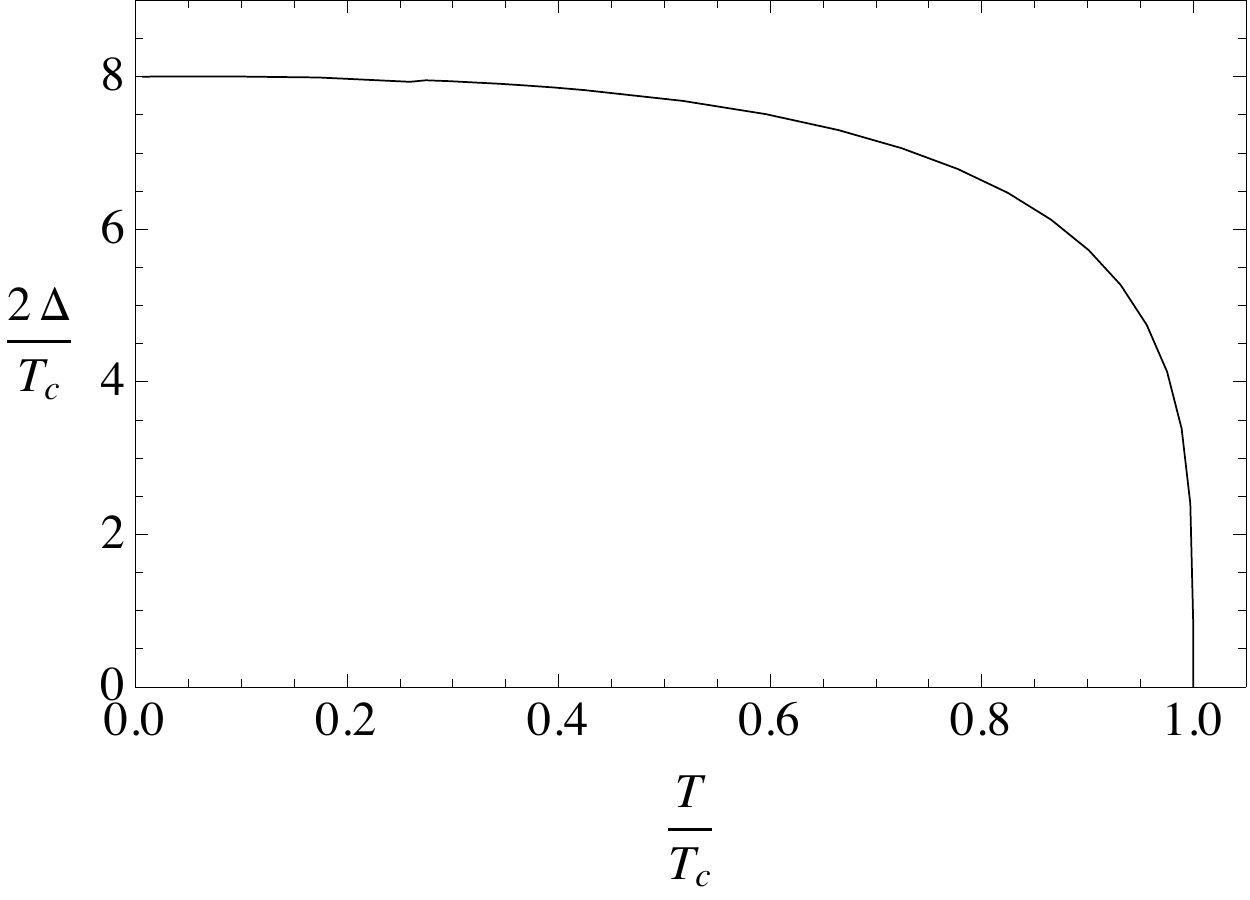}
\caption{The gap as a function of temperature.}\label{order}
\end{center}\end{figure}

If we fix the charge density $\rho$, one finds that the hairy black holes
only exist below a critical temperature
\be
T_c = 0.125 \sqrt{\rho} \,.
\ee
See figure \ref{order}. With some hindsight, see the following section, we introduce the
gap notation
\be
2 \Delta = \sqrt{\langle J \rangle} \,.
\ee

Note that $J$ is an operator with mass dimension 2, appropriate
for a current density in $2+1$ dimensions. The natural
dimensionless quantities to plot are therefore $2 \Delta/T_c$
against $T/T_c$. Just below the critical temperature we find
\be
\langle J \rangle \approx 104.8 T_c^2 (1- T/T_c)^{1/2} \,,
\ee
as expected for a mean field second order transition. It is interesting to
note that the zero temperature gap is
\be
\frac{2 \Delta(0)}{T_c} \approx 8 \,.
\ee
This is greater than the BCS value, rather close to the value
found in the holographic model of \cite{Hartnoll:2008vx} and
consistent with some values observed in the cuprates. Given the
similarity of this value for the gap and the one obtained in
\cite{Hartnoll:2008vx}, it would be of interest to obtain values
for different holographic models to see to what extent it is
universal.

\section{Conductivity and spectral density}

Two basic quantities characterising any superconductor are the
frequency dependent conductivity and the spectral density. These
are closely related, but not identical in the presence of a Hall
conductivity. The electrical conductivity is defined through Ohm's
law
\be\label{eq:ohm}
J_i = \sigma_{ij} E^j \,.
\ee
Here $E^j$ is an external electric field and $J_i$ the current
generated. Because the superconducting phase we are studying is
isotropic, we will have $\sigma_{xx} = \sigma_{yy}$ and
$\sigma_{xy} = - \sigma_{yx}$. In a non-isotropic phase the distinction
between the standard and Hall conductivites is not well defined.

We will compute the conductivity directly from Ohm's law
(\ref{eq:ohm}). However, to relate the conductivity to the
spectral density, we note the standard formula from linear
response theory
\be\label{eq:kubo}
\sigma_{ij}(\w) = \frac{- i G_{ij}^R(\w)}{\w} \,,
\ee
in which the retarded Greens function is
\be\label{eq:green}
G_{ij}^R(\w) = - i \int d^2x dt e^{-i \w t} \theta(t) \langle
[J_i(t), J_j(0)] \rangle \,.
\ee
The spectral density is defined to be twice the imaginary part of
the retarded Greens function. However, in order to obtain a
sensible, i.e. non-negative, spectral density one needs the Greens
function of currents that do not couple. These are obtained by
diagonalising the conductivity. The conductivity matrix has
eigenvalues
\be\label{eq:szszb}
\sigma_z = \sigma_{xx} + i \sigma_{xy} \,, \qquad \sigma_{\bar z} = \sigma_{xx} - i
\sigma_{xy}\,.
\ee
In the eigenbasis Ohm's law becomes the decoupled equations $J_z =
\sigma_z E_z$ and $J_{\bar z} = \sigma_{\bar z} E_{\bar z}$. We
are using the complex coordinates introduced in (\ref{eq:z})
above. This complexified formalism is similar to that used in
\cite{Hartnoll:2007ip}. The spectral densities associated with the
decoupled currents $J_z$ and $J_{\bar z}$ are thus
\be\label{eq:spectral}
\chi_z(\w) = 2 \w \text{Re}\, \sigma_z(\w) \,, \qquad \chi_{\zb} (\w) =
2\w \text{Re}\, \sigma_{\zb} (\w)\,.
\ee
We recall that the physical interpretation of the spectral density
is that it gives us the density of energy eigenstates at energy
$\w$, weighted by their overlap with the electric current
operators. Therefore the spectral density is the appropriate
quantity with which to probe the pseudogap region, that will make
an appearance shortly.

In order to compute linear response functions, such as the
conductivity, using the gravitational dual to the superconductor,
one needs to consider linearised perturbations of the fields about
the black hole background \cite{Son:2002sd}. Specifically, we are
interested in the $\t^3$ component of the nonabelian current,
$j^3_{x,y}$. This will be dual to fluctuations of the $A^3_{x,y}$
fields. Because of the nonlinearities in the Yang-Mills action,
fluctuations in $A^3_{x,y}$ will source other fields. We need to
keep all the modes that are coupled at a linearised level for
consistency.

Since our background is invariant under a combined spatial and
gauge rotation (\ref{eq:combined}), it is sufficient to consider
fields that have the same charge as $A^3_{x,y}$ under this $U(1)$
action. Specifically, we see that there will be decoupled
equations involving the sets of fields
\be
\{A_r^+(r), A_t^+(r), A_z^3(r) \} \qquad \text{and} \qquad \{A_r^-(r), A_t^-(r),
A_{\zb}^3(r) \}\,.
\ee
All of these fields are taken to have an overall time dependence
of $e^{-i\omega t}$.

There is still some gauge freedom left, which reduces the actual
degrees of freedom. In particular, we can consider a background
field gauge transformation generated by some $\lambda^a$
\be
\delta_{BG}A^a_\mu = \partial_\mu \lambda^a+
f^{abc} A^b_\mu \lambda^c\,.
\ee
We will use this freedom to set $A_r^\pm = 0$. The linearized
equations for the perturbations become
\bea \label{num1}
{A^3_z}''+\left(\frac{2}{r}+\frac{h'}{h}\right){A_z^3}'+\frac{\omega^2-hw^2}{r^4h^2}{A_z^3}-\frac{w(\phi+\omega)}{r^4h^2}{A_t^+}
& = & 0
\,, \\
\label{num2}
{A^3_{\zb}}''+\left(\frac{2}{r}+\frac{h'}{h}\right){A^3_{\zb}}'+\frac{\omega^2-hw^2}{r^4h^2}A^3_{\zb}-\frac{w(\phi-\omega)}{r^4h^2}{A_t^-}
& = & 0
\,, \\
\label{red1}
{A_t^+}''+\frac{2}{r}{A_t^+}'-\frac{w^2}{r^4h}
{A_t^+}+\frac{w(\phi+\omega)}{r^4h}{A_z^3} & = & 0
\,, \\
\label{red2}
{A_t^-}''+\frac{2}{r}{A_t^-}'-\frac{w^2}{r^4h}
{A_t^-}+\frac{w(\phi-\omega)}{r^4h}{A_{\zb}^3} & = & 0
\,, \\
\label{num3}
{A_t^+}\phi'+h(w'{A_z^3}-w{A_z^3}')-(\phi-\omega){A_t^+}'& = & 0
\,, \\
\label{num4}
{A_t^+}\phi'+h(w'{A_z^3}-w{A_{\zb}^3}')-(\phi+\omega){A_t^-}' & =
&0
\,.
\eea
Two of these equations are redundant: Equations
\eqref{red1} and \eqref{red2} can be derived by differentiating the
first order equation and substituting the other equations of
motion. Therefore, in solving these equations numerically it is
sufficient to numerically integrate
\eqref{num1},~\eqref{num2},~\eqref{num3} and \eqref{num4}. Despite
the first-derivative equations appearing to have singular points,
they numerically integrate perfectly well.

The equations are solved numerically by integrating out from the
horizon to infinity. In order to obtain retarded Greens functions,
ingoing boundary conditions must be imposed at the horizon
\cite{Son:2002sd}. This is most conveniently done by requiring the
following behaviour near the horizon $r \approx 1$:
\bea
A_{z}^3 & = & h^{-i\omega/3}a_0+ \cdots\,, \\
A_{\zb}^3 & = & h^{-i\omega/3}b_0+ \cdots \,, \\
A_{t}^+ & = & \frac{a_0~w_0}{i+\omega/3}h^{-i\omega/3}(r-1)+ \cdots \,, \\
A_{t}^-& = & - \frac{b_0~w_0}{i+\omega/3}h^{-i\omega/3}(r-1)+ \cdots \,.
\eea
Recall that we defined $w_0$ above as the value of the background
field $w$ at the horizon. Given the background, there are two free
constants, $a_0$ and $b_0$.

Integrating the fields out to large $r$, we can read off the dual currents and external
electric fields. The current and charge densities are obtained from
\be\label{eq:F1}
F^a_{r\mu} = \frac{g^2 \langle J^a_\mu \rangle}{r^2} + \cdots \,,
\ee
where $\mu$ here runs over the boundary directions $t, z, \zb$. In this expression
we have included the Yang-Mills coupling $g$ due to the action (\ref{eq:action}).
This coupling determines the (constant) conductivity of the normal, non-superconducting,
state \cite{Herzog:2007ij}
\be
\sigma_n = \frac{1}{g^2} \,.
\ee
The numerical value of the normal state conductivity depends on the theory.
The external electric fields then are obtained from
\be\label{eq:F2}
F^a_{t i} = - E^a_i + \cdots \,.
\ee
(A more symmetric formulation of (\ref{eq:F1}) and (\ref{eq:F2})
is possible in terms of the radial variable $u = 1/r$.) From the
previous section, we have the background equilibrium values
\be
\langle J^\pm_{z, \zb} \rangle = - J \,, \qquad \langle J^3_{t} \rangle = \rho \,.
\ee

In reading off the linearised electric response to a time varying
external field we have to face the fact that the na\"ive $SU(2)$
conductivity, or the projection of it onto the $\t^3$ direction,
is not $SU(2)$ invariant. We are interested in the electrical
conductivity of the $U(1)$ subgroup of $SU(2)$ generated by
$\t^3$. Therefore, we should consider currents that result from
external sources in the $\t^3$ direction only. In many
circumstances it would be sufficient to ensure that we have
electric field $E_i \equiv E_i^3$ only. However, because the
na\"ive rotational invariance is broken, there is a nontrivial
charge-current density Green's function $G_{it}^R(\w)$. Therefore,
we need a configuration in which the background $A_t^\pm$ also
vanishes. Here $A_t^\pm$ is a source for the charge density
$\rho^\pm$.

If we were to simply take our equations \eqref{num1} to
\eqref{num4} and integrate them to the boundary, while we would
never obtain electric fields $E^\pm_i$, we would obtain a source
$A_t^\pm$. We therefore need to do an $SU(2)$ rotation to set this
term to zero. In the bulk this means we should allow for a gauge
transformation that sets the boundary value of $A_t^\pm$ to zero.
A gauge transformation generated by $\lambda^a(r)$ results in the
new scalar potential
\be
\delta A_t^\pm = A_t^\pm - i \lambda^\pm (\w \mp \phi) \,,
\ee
as well as the new field strengths
\bea
\delta F^\pm_{rt} & = & {A_t^\pm}' \pm i\lambda^\pm \phi'  \,, \\
\delta F^3_{rz} & =  & {A^3_z}'- i\lambda^+ w' \,, \\
\delta F^3_{r\zb} & =  & {A^3_{\zb}}'+ i\lambda^- w' \,, \\
\delta F^3_{tz} & = & i \w A_z^3 + w ( i A_t^+ \mp \lambda^+ \phi)
\,.
\eea
In the final of these expressions, only the first term contributes
at the boundary $r \to \infty$.

We wish to cancel the leading asymptotic term in $A^\pm_{t}$ with
$\lambda$. After doing this, we can obtain the conductivities
\bea\label{eq:zconduct}
\sigma_{z} & = & \frac{J_{z}}{E_{z}} = - \sigma_n \lim_{r \to \infty} \frac{r^2 \delta F_{rz}^3}{\delta F_{tz}^3}
= \sigma_n \lim_{r \to \infty} \frac{i}{\w A_{z}^3}
\left(r^2 A_z^3{}'  + \frac{J}{\sqrt{2} (\w - \mu)} A_{t}^+ \right) \,, \\
\sigma_{\zb} & = & \frac{J_{\zb}}{E_{\zb}} = - \sigma_n \lim_{r \to \infty} \frac{r^2 \delta F_{r\zb}^3}{\delta F_{t\zb}^3}
= \sigma_n \lim_{r \to \infty} \frac{i}{\w A_{\zb}^3}
\left(r^2 A_{\zb}^3{}' - \frac{J}{\sqrt{2} (\w + \mu)} A_{t}^- \right)
\,.
\eea
We can see that this expression has a curious pole at $\w = \mu$.
This will appear as a delta function in the spectral density for
$J_{z}$ \footnote{We thank Gubser and Pufu for convincing us
that this pole is indeed physical.}. It is of interest to
elucidate the physics behind this stable resonance, but we shall
not do so here.

\begin{figure}\begin{center}
\includegraphics[width=.45\textwidth]{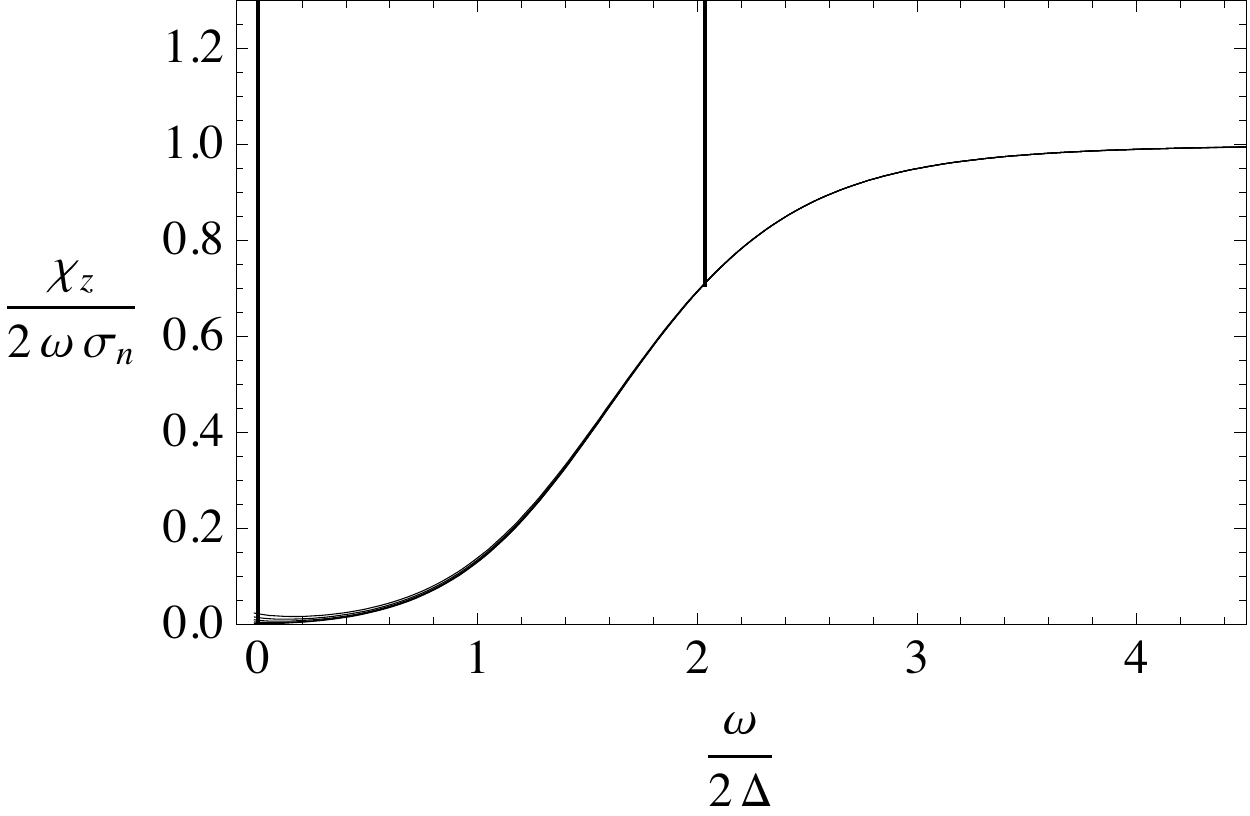}\hspace{0.2cm}
\includegraphics[width=.45\textwidth]{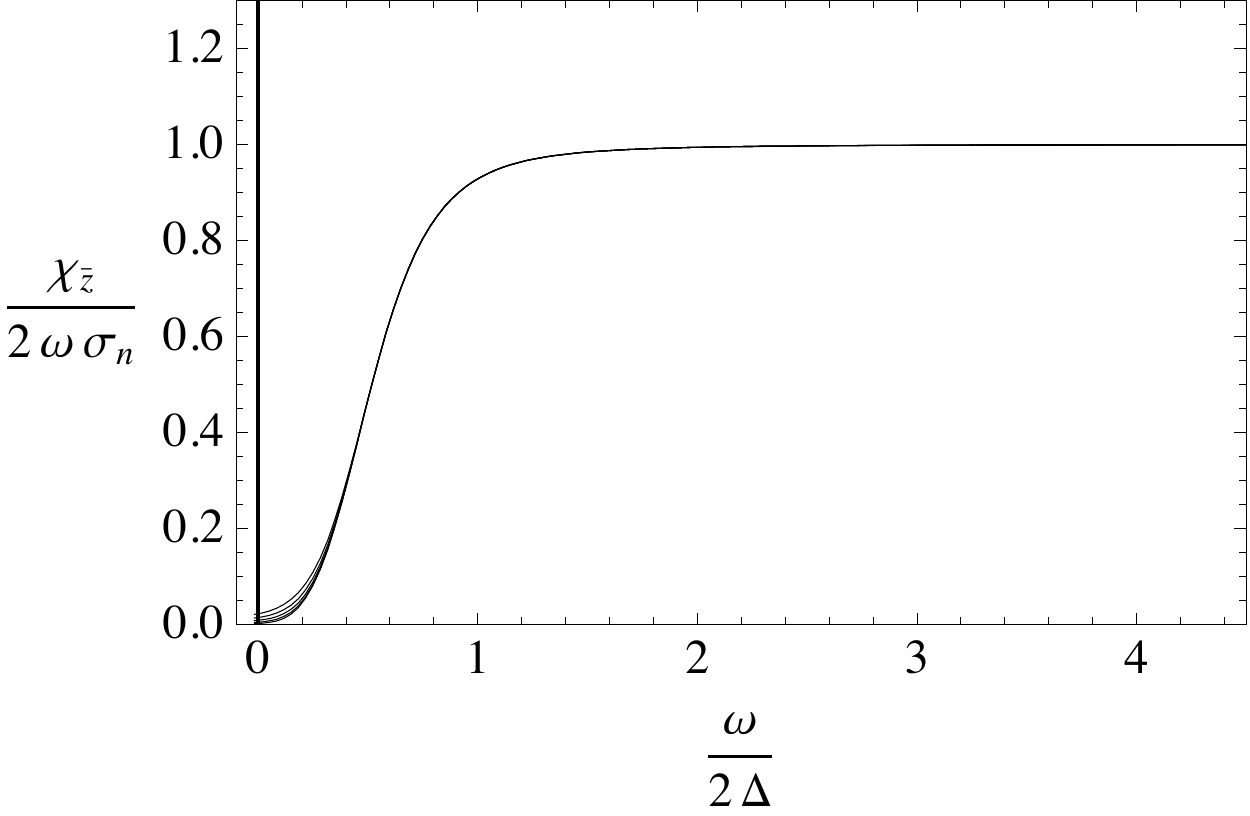}
\caption{Spectral functions for the currents $J_z$ and $J_{\zb}$. Each plot is in fact five curves,
at temperatures $T/T_c=.08,.11,.15,.19,.23$. We see that we have
effectively reached the zero temperature limit. There is a clear
pseudogap, and a delta function at $\w=0$. There is also a delta
function at $\w = \mu$.}\label{sigmaz}
\end{center}\end{figure}

\begin{figure}\begin{center}
\includegraphics[width=0.45\textwidth]{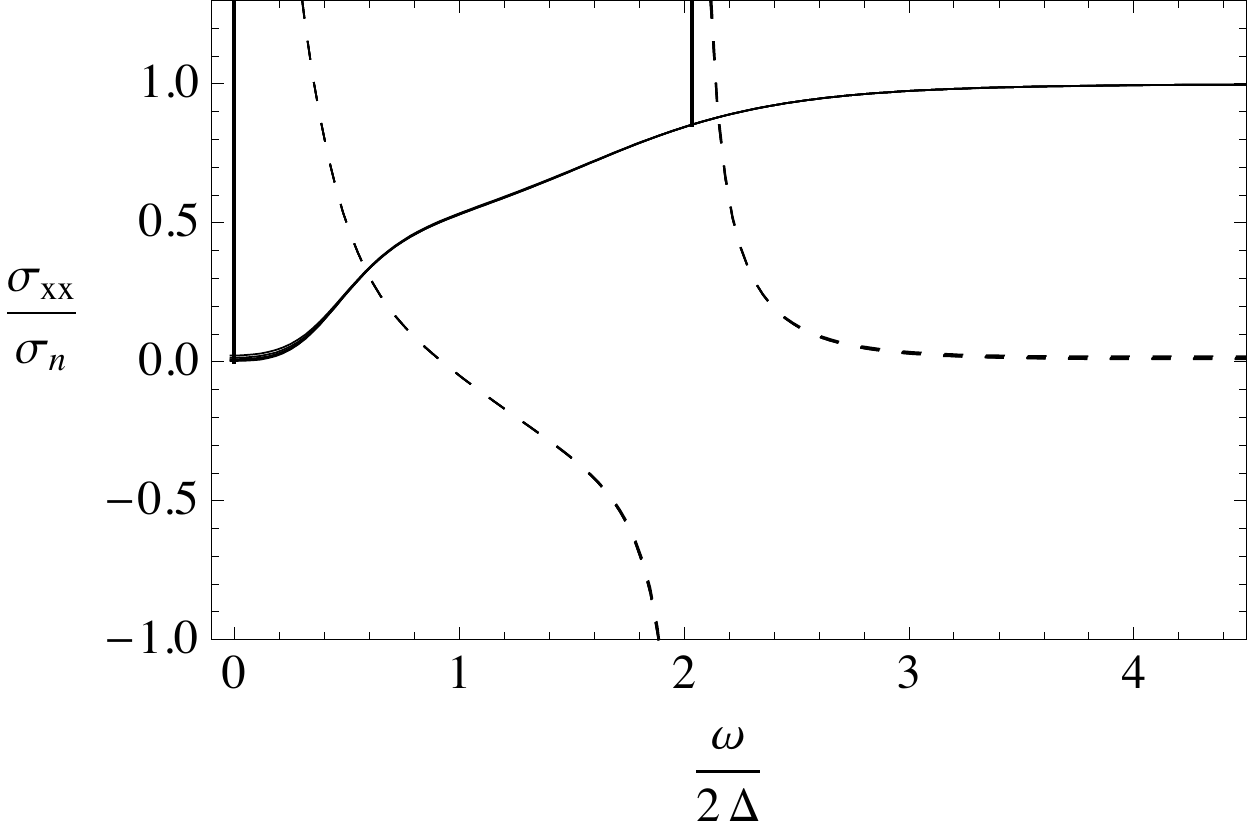}\hspace{0.2cm}
\includegraphics[width=0.45\textwidth]{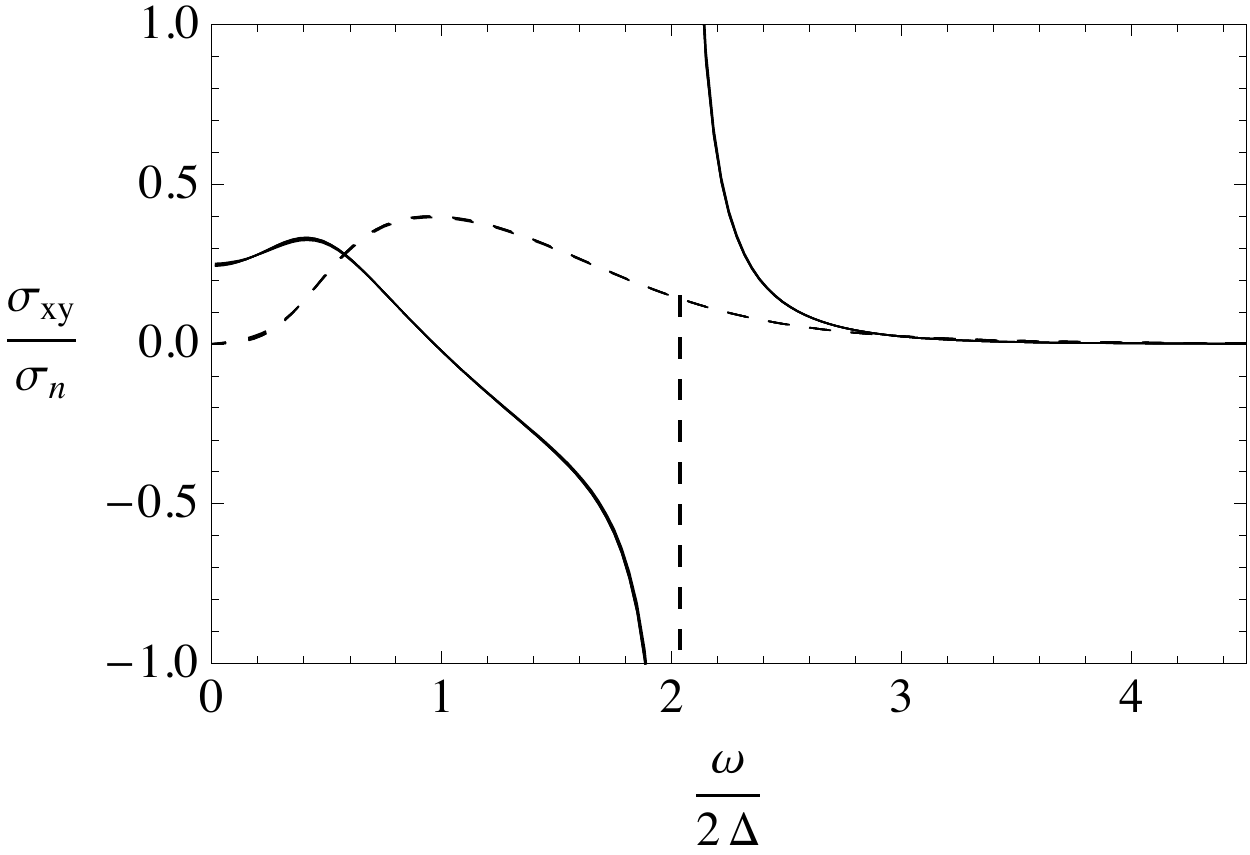}
\caption{Standard and Hall
conductivities at low temperatures as a function of frequency. The
solid lines are the real part whereas the dashed lines are the
imaginary parts. A pole in the imaginary part at $\w = 0$
indicates that the real part will contain a delta function at $\w
= 0$. Similarly with the pole at $\w = \mu$}\label{sigmaxx}
\end{center}\end{figure}

Given $\sigma_z$ and $\sigma_{\zb}$ we can obtain the spectral
functions directly from (\ref{eq:spectral}) and the standard and
Hall conductivities by inverting (\ref{eq:szszb}) to obtain
\be
\sigma_{xx}=  \frac{1}{2} \left(\sigma_z+\sigma_{\zb} \right) \,, \qquad
\sigma_{xy}= \frac{-i}{2} \left(\sigma_z-\sigma_{\zb} \right) \,.
\ee
Our result for the spectral functions is plotted in figure
\ref{sigmaz}. The standard and Hall conductivities are plotted in
figure \ref{sigmaxx}.

The principal qualitative features of our results are immediately
seen in these plots. Firstly, there is a strong depletion of
spectral weight at small frequencies. However, even as we take the
temperature to zero, the spectral function does not go to zero at
these low frequencies. Therefore, we will call this region the
pseudogap. The two currents, $J_z$ and $J_{\zb}$ have pseudogaps
of differing widths. We have defined $2 \Delta$ to be the width of
the narrower of the two, see the right hand plot of figure
\ref{sigmaz}. Obviously there is some ambiguity in the definition
of the width. The two different gaps are likely due to the fact
that there are effectively two condensates in the ansatz
(\ref{eq:ansatz}). It would be interesting to disentangle their
physics by considering more general backgrounds with more than one
radial function. Allowing for gauge transformations,
(\ref{eq:ansatz}) can be generalised to three independent radial
functions. The only isotropic ansatz is (\ref{eq:ansatz}), up to
gauge transformations.

In the standard conductivity we find a delta function in the real
part at $\w=0$. This is the Goldstone boson of the spontaneously
broken $U(1)$ symmetry and is the signal of superconductivity. We
again see a pseudogap at low frequencies. Because this
conductivity is sensitive to both the $\chi_z$ and $\chi_{\zb}$
spectral functions, which had gaps of differing widths, there is a
small feature in the conductivity.

We also find a nonvanishing Hall conductivity. This Hall
conductivity is different to previous appearances of Hall
conductivity in AdS/CFT, e.g. \cite{Hartnoll:2007ai}, in that it
is not due to an external magnetic field. Rather, a Hall
conductivity is possible because the superconducting condensate
broke time reversal invariance. This can be seen directly from
(\ref{eq:kubo}) and (\ref{eq:green}). Recalling that T is an
antilinear operator, then (\ref{eq:green}) implies that
$\sigma_{ij}(\w) = \sigma_{ji}(\w)$ in a state that is
T-invariant\footnote{$\langle [J_i(t), J_j(0)] \rangle = \langle T
[J_i(t), J_j(0)]T \rangle^* = \langle [J_j(0), J_i(-t)] \rangle =
\langle [J_j(t), J_i(0)] \rangle$.}. Isotropy implies $\sigma_{xy}(\w) = -
\sigma_{yx}(\w)$, and therefore T-invariance implies
$\sigma_{xy}(\w)=0$. Furthermore (recall that the conductivity is
dimensionless in 2+1 dimensions) we find that the Hall
conductivity attains a finite real value at $\w = 0$.

Now let us extract some quantitative results from our data. The superfluid density is the coefficient of the delta function in the real conductivity at $\w = 0$. By the Kramers-Kronig relations, this is also the coefficient of the pole in the imaginary part of the conductivity as $\w \to 0$:
\be
\mathrm{Re}\, \sigma_{xx} (\omega) \sim \pi n_s \delta(\omega)\quad \rightarrow \quad \mathrm{Im}\,\sigma_{xx}(\omega)\sim n_s/\omega \,.
\ee
From our equations we find
\bea
\text{for} \quad T \ll T_c: & \quad n_s & \approx  \quad C \sigma_n \Delta \,, \\
\text{as} \quad T \to T_c: & \quad n_s & \approx \quad C'\sigma_n (T_c-T) \,,
\eea
where the numerical coefficients are $C=0.25,~C'=12.2.$
As in \cite{Hartnoll:2008vx}, we can note that a linearly vanishing superfluid density near the critical
temperature results in a London magnetic penetration depth of $\lambda_L \sim (T_c-T)^{-1/2}$,
as expected from Landau-Ginzburg theory.

We can also read off the normal component of the DC superconductivity. In contrast
to the fully gapped model of \cite{Hartnoll:2008vx}, there is no exponential suppression
at small temperatures. Instead we find a quadratic dependence on temperature
\be
n_n = \lim_{\omega\rightarrow 0}\mathrm{Re}\, \sigma_{xx} \approx
0.32
\, \sigma_n (T/T_c)^2 + \cdots \,,
\ee
and in the zero temperature limit  the small frequency dependence is
\be
\mathrm{Re}\, \sigma_{xx} \approx
0.20
\, \sigma_n (\omega/2\Delta)^2 + \cdots \,.
\ee
Similarly, we can examine the DC hall effect and find
\be
H = \lim_{\omega\rightarrow0}\mathrm{Re}\, \sigma_{xy} \approx  0.24 \sigma_n
+ 0.13 \sigma_n (T/T_c)^2 + \cdots \,.
\ee
There is no superfluid component because the imaginary part has no pole. On the other hand,
because there is no Hall conductivity in the normal phase, it is clear that superconducting physics
is playing an important role.

The crucial question is of course to understand the physics
underlying the pseudogap in this model. One might hope that the
psuedogap is indeed a signal of the $p$- or $d$-wave nature of the
`Cooper pairs'. The operator that condenses is a component of a
global $SU(2)$ current. In many cases this will be a bilinear in
UV operators, including fermions. The combination of spacetime and
internal symmetries that occurs in these models to preserve
isotropy is also reminiscent of non $s-$wave superconductors. In
order to be completely sure that there is indeed a non $s-$wave
condensate one should probe the system as a function of momentum
$k_i$ and look for a momentum-dependent gap. We hope to address
this in the future.

Another possibility is that the pseudogap is directly due to the
fact that there is a massless excitation in the theory: the
Goldstone boson\footnote{We'd like to thank Dam Son for drawing
this possibility to our attention.}. The electric current can
decay at arbitrarily small energy into a multi Goldstone boson
state. By the optical theorem, this process should introduce a
branch cut in the current-current Greens function reaching down to
$\w=0$. In fact, one might ask why such a cut didn't show up in
the conductivity computations in \cite{Hartnoll:2008vx}. At weak
coupling such processes are at higher loop order and are
suppressed relative to pair production of `electrons' (the
quasiparticles forming the Cooper pair). At strong coupling it is
less clear. Perhaps such cuts reaching $\w=0$ are suppressed at
large $N$ in a similar manner to those resulting in hydrodynamic
tails \cite{Kovtun:2003vj}. Alternatively, perhaps breaking $T$
invariance allows decay channels that were not possible in
\cite{Hartnoll:2008vx}.

It is certainly of interest to look for possible embeddings of
Yang-Mills theory in AdS into string or M theory, ideally
consistent with the probe approximation. A natural way to get an
$SU(2)$ field in string theory is by using coincident D branes. In
such a setup the probe limit will be admissible at weak string
coupling (large $N$). For instance, the supersymmetric D3-D5
system, with $N$ D3 branes and 2 D5 branes, admits a near horizon
description as two probe D5 branes in $AdS_5 \times S^5$. The
probe branes lie on an $AdS_4$ in $AdS_5$ and have an $SU(2)$
gauge field on their worldvolume, thus precisely realising our
setup. Although the remaining directions of the D5 branes form an
$S^2$ in $S^5$, they are stable because the mass of the slipping
mode is above the Breitenlohner-Freedman bound.

The Lagrangian (\ref{eq:action}) can also be uplifted to eleven
dimensional supergravity on $AdS_4 \times S^7$ using
\cite{Mann:2006jc}, following \cite{Cvetic:1999au}. However, this
lift gives $g^2 L^2 = \kappa^2_4/4$ (at least according to
\cite{Mann:2006jc}) \footnote{In fact \cite{Mann:2006jc} and \cite{Cvetic:1999au}
differ by a factor of 2 for $g^2 L^2$. Both values are below the
critical value for superconductivity. We thank Gubser and Pufu for
bringing these facts to our attention.}. This appears to be below
the critical value $g^2 L^2 \approx \kappa_4^2$ needed for a
superconducting instability to occur \cite{Gubser:2008zu}.
Therefore, this lift does not realise the physics of interest.

\section{Discussion and stability}
\label{sec:stability}

It was shown in  \cite{Gubser:2008wv}  that, at least near to the
critical temperature, the isotropic superconducting phase is
unstable to a perturbation breaking rotational invariance. This
was achieved by searching for dynamical instabilities, that is,
normalisable modes that grow exponentially in time. Their
postulated endpoint of this instability is an anisotropic phase
with background ansatz
\be\label{eq:ansatz2}
A=\phi(r)~dt\tau^3+ w(r)~dx \tau^1 \,.
\ee
This background breaks time reversal symmetry, as well as being
incurably anisotropic. The anisotropy means that it is not
possible to invariantly separate the standard and Hall
conductivities. Therefore one should look for different signals of
the time reversal breaking.

To confirm this instability and follow it down to lower
temperatures, we have computed the free energies of the isotropic
and anisotropic phases. The two backgrounds come into existence at
the same critical temperature $T_c$. We have been able to study
the backgrounds down to $T/T_c\approx 0.23$. Working in the grand
canonical ensemble, that is, fixed chemical potential, to compute
the free energy it is sufficient to simply compare the Yang-Mills
action (\ref{eq:action}) evaluated on the solutions, with the same
chemical potential (and not necessarily the same charges).

\begin{figure}[h]
\begin{center}
\includegraphics[width=3in]{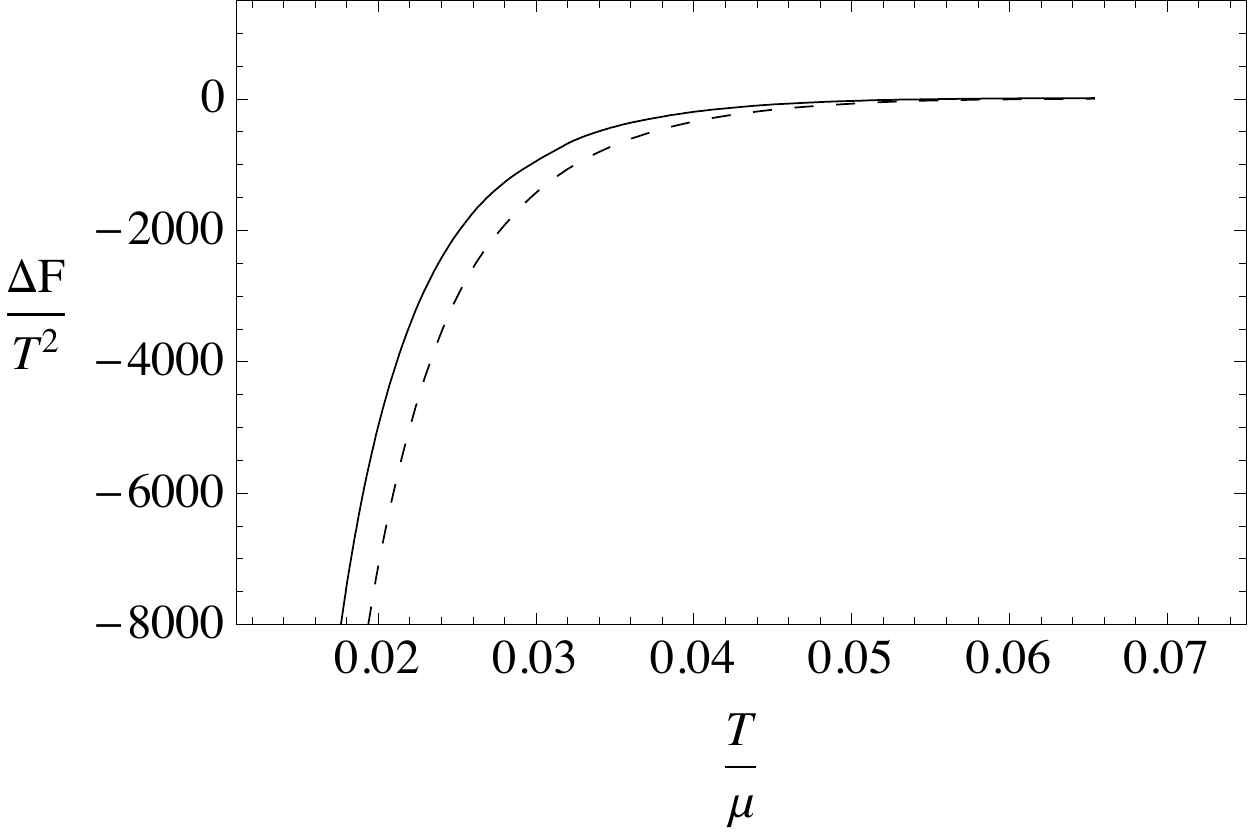}
\caption{Comparison of free energies for temperatures below $T_c$. The anisotropic phase is a dashed line. It always has a lower free energy than the isotropic phase.}\label{instability}
\end{center}
\end{figure}

The result is shown in figure \ref{instability}.  For each phase
we have plotted the difference in free energy compared to the
phase without a condensate. We see that the anisotropic phase
appears to be favoured at all temperatures.

The mode about the isotropic black hole leading to an instability
does not mix with the modes we have been considering. Therefore
the electrical response of the theory remains well defined at the
level of linear response. The phase we have studied is interesting
therefore as a model of a strongly coupled isotropic
superconductor with a pseudogap and spontaneous time reversal
breaking. These are four properties (strongly coupled, isotropic,
pseudogapped and T-non-invariant) that are shared by
superconductors of experimental interest. It would of course be
interesting to identify other experimental quantities that are
amenable to computation in this model.

Further on the question of instability, one should note that there
are other backgrounds than (\ref{eq:ansatz}) and
(\ref{eq:ansatz2}) that are possible, and which are natural
candidates for the dominant state. After allowing for gauge
transformations it seems that the general ansatz has three
independent radial functions. This more general ansatz deserves
study.

The main results of this paper have been to exhibit a pseudogap
and a Hall conductivity in an isotropic phase of the holographic
superconductor proposed in \cite{Gubser:2008zu}. A primary
question for future work is to explain the pseudogap in this
model. We noted that two natural candidates are firstly non
$s-$wave pairing and secondly intermediate states of massless
Goldstone bosons.

A second question is to find upliftings of the Yang-Mills model
into string and M theory. Probe D branes are a natural way to
engineer nonabelian symmetries and will furthermore be consistent
with the probe limit.

Finally, we observed that the zero temperature gap was $2
\Delta/T_c \approx 8$. This is very close to the value found in the
holographic superconductor studied in \cite{Hartnoll:2008vx}. It
would be interesting to study more examples and see to what extent
this value is universal for holographic superconductors.

\section*{Acknowledgements}

We would like to thank Gary Horowitz for many helpful comments. It
is also a pleasure to acknowledge discussions with Leon Balents,
Matthew Fisher, David Mateos, Rafael Porto and Dam Son. We are
grateful to Gubser and Pufu for their comments on the first
preprint version of this paper. This research was supported in
part by the National Science Foundation under Grants No.
PHY05-51164 and PHY-0555669.


\begin{thebibliography}{99}

\bibitem{Bardeen:1957mv}
  J.~Bardeen, L.~N.~Cooper and J.~R.~Schrieffer,
  ``Theory Of Superconductivity,''
  Phys.\ Rev.\  {\bf 108}, 1175 (1957).

\bibitem{parks}
  R.~D.~Parks, \textit{Superconductivity}, Marcel Dekker Inc.
  (1969).

 \bibitem{heavy}
P.~Gegenwart, Q.~Si and F.~Steglich, ``Quantum criticality in
heavy-fermion metals,'' Nature Physics, {\bf 4} (2008) 186.

\bibitem{heavy2}
P.~Monthoux, D.~Pines and G.~G.~Lonzarich, ``Superconductivity
without phonons,'' Nature, {\bf 450} (2008) 1177.

\bibitem{hightc}
  E.~W.~Carlson, V.~J.~Emery, S.~A.~Kivelson and D.~Orgad,
  ``Concepts in high temperature superconductivity,''
  arXiv:cond-mat/0206217.

\bibitem{Maldacena:1997re}
J.~M.~Maldacena, ``The large N limit of superconformal field
theories and supergravity,'' Adv.\ Theor.\ Math.\ Phys.\  {\bf 2}
(1998) 231 [Int.\ J.\ Theor.\ Phys.\  {\bf 38} (1999) 1113]
[arXiv:hep-th/9711200].

\bibitem{Gubser:2005ih}
  S.~S.~Gubser,
  ``Phase transitions near black hole horizons,''
  Class.\ Quant.\ Grav.\  {\bf 22}, 5121 (2005)
  [arXiv:hep-th/0505189].

\bibitem{Gubser:2008px}
  S.~S.~Gubser,
  ``Breaking an Abelian gauge symmetry near a black hole horizon,''
  arXiv:0801.2977 [hep-th].

\bibitem{Hartnoll:2008vx}
  S.~A.~Hartnoll, C.~P.~Herzog and G.~T.~Horowitz,
  ``Building an AdS/CFT superconductor,''
  arXiv:0803.3295 [hep-th].

\bibitem{Gubser:2008zu}
  S.~S.~Gubser,
  ``Colorful horizons with charge in anti-de Sitter space,''
  arXiv:0803.3483 [hep-th].

\bibitem{Herzog:2007ij}
  C.~P.~Herzog, P.~Kovtun, S.~Sachdev and D.~T.~Son,
  ``Quantum critical transport, duality, and M-theory,''
  Phys.\ Rev.\  D {\bf 75}, 085020 (2007)
  [arXiv:hep-th/0701036].

\bibitem{Hartnoll:2007ih}
  S.~A.~Hartnoll, P.~K.~Kovtun, M.~Muller and S.~Sachdev,
  ``Theory of the Nernst effect near quantum phase transitions in condensed
  matter, and in dyonic black holes,''
  Phys.\ Rev.\  B {\bf 76}, 144502 (2007)
  [arXiv:0706.3215 [cond-mat.str-el]].

\bibitem{Hartnoll:2007ip}
  S.~A.~Hartnoll and C.~P.~Herzog,
  ``Ohm's Law at strong coupling: S duality and the cyclotron resonance,''
  Phys.\ Rev.\  D {\bf 76}, 106012 (2007)
  [arXiv:0706.3228 [hep-th]].

\bibitem{Hartnoll:2008hs}
  S.~A.~Hartnoll and C.~P.~Herzog,
  ``Impure AdS/CFT,''
  arXiv:0801.1693 [hep-th].

\bibitem{Son:2008ye}
  D.~T.~Son,
  ``Toward an AdS/cold atoms correspondence: a geometric realization of the
  Schroedinger symmetry,''
  arXiv:0804.3972 [hep-th].

\bibitem{Balasubramanian:2008dm}
  K.~Balasubramanian and J.~McGreevy,
  ``Gravity duals for non-relativistic CFTs,''
  arXiv:0804.4053 [hep-th].

\bibitem{sachdev}
S. Sachdev, {\it Quantum Phase Transitions}, CUP, 1999.

\bibitem{one} M.~Covington, M.~Aprili, E.~Paraoanu, L.~H.~Greene,
F.~Xu, J.~Zhu and C.~A.~Mirkin, ``Observation of surface-induced
broken time-reversal symmetry in YBa$_2$Cu$_3$O$_7$ tunnel
junctions,'' Phys. Rev. Lett. {\bf 79} (1997) 277.

\bibitem{two} R.~Krupke and G.~Deutscher, ``Anisotropic magnetic
field dependence of the zero-bias anomaly on in-plane oriented
[100] Y$_1$Ba$_2$Cu$_3$O$_{7-x}$/In tunnel juctions,'' Phys. Rev.
Lett. {\bf 83} (1999) 4634.

\bibitem{three} R.~Carmi, E.~Polturak, G.~Koren and A.~Auerbach,
``Spontaneous macroscopic magnetization at the superconducting
transition temperature of YBa$_2$Cu$_3$O$_{7-\delta}$,'' Nature
{\bf 404} (2000) 853.

\bibitem{four} G.~Deutscher, Y.~Dagan, A.~Kohen and R.~Krupke,
``Field induced and spontaneous sub-gap in [110] and [100]
oriented YBCO films: indication for a $d_{x^2-y^2}+i d_{xy}$ order
parameter,'' Physica C {\bf 341-348} (2000) 1629.

\bibitem{five} Y.~Dagan and G.~Deutscher, ``On the origin of time
reversal symmetry breaking in Y$_{1-y}$Ca$_y$ Ba$_2$ Cu$_3$
O$_7-x$,'' Europhys. Lett. {\bf 57} (2002) 444.

\bibitem{tone} R.~B.~Laughlin, ``Magnetic induction of $d_{x^2-y^2}+i d_{xy}$
order in high-$T_c$ superconductors,'' Phys. Rev. Lett. {\bf 80}
(1998) 5188.

\bibitem{ttwo} J.~Goryo and K.~Ishikawa, ``Observation of induced
Chern-Simons term in P- and T- violating superconductors,'' Phys.
Lett. {\bf A260} (1999) 294.

\bibitem{Gubser:2008wv}
  S.~S.~Gubser and S.~S.~Pufu,
  ``The gravity dual of a p-wave superconductor,''
  arXiv:0805.2960 [hep-th].

\bibitem{Karch:2007pd}
  A.~Karch and A.~O'Bannon,
  ``Metallic AdS/CFT,''
  JHEP {\bf 0709}, 024 (2007)
  [arXiv:0705.3870 [hep-th]].

\bibitem{Son:2002sd}
  D.~T.~Son and A.~O.~Starinets,
  ``Minkowski-space correlators in AdS/CFT correspondence: Recipe and
  applications,''
  JHEP {\bf 0209}, 042 (2002)
  [arXiv:hep-th/0205051].

\bibitem{Hartnoll:2007ai}
  S.~A.~Hartnoll and P.~Kovtun,
  ``Hall conductivity from dyonic black holes,''
  Phys.\ Rev.\  D {\bf 76}, 066001 (2007)
  [arXiv:0704.1160 [hep-th]].

\bibitem{Kovtun:2003vj}
  P.~Kovtun and L.~G.~Yaffe,
  ``Hydrodynamic fluctuations, long-time tails, and supersymmetry,''
  Phys.\ Rev.\  D {\bf 68}, 025007 (2003)
  [arXiv:hep-th/0303010].

\bibitem{Mann:2006jc}
  R.~B.~Mann, E.~Radu and D.~H.~Tchrakian,
  ``Nonabelian solutions in AdS(4) and d = 11 supergravity,''
  Phys.\ Rev.\  D {\bf 74}, 064015 (2006)
  [arXiv:hep-th/0606004].

\bibitem{Cvetic:1999au}
  M.~Cvetic, H.~Lu and C.~N.~Pope,
  ``Four-dimensional N = 4, SO(4) gauged supergravity from D = 11,''
  Nucl.\ Phys.\  B {\bf 574}, 761 (2000)
  [arXiv:hep-th/9910252].

\end{thebibliography}
\end{document}